  \documentclass[10pt,aps,prl,groupedaddress,floatfix,footinbib]{revtex4-2}

\usepackage{etoolbox}
\usepackage{graphicx}
\usepackage{dcolumn}
\usepackage{bm}

\usepackage{graphicx}%
\usepackage{multirow}%
\usepackage{amsmath,amssymb,amsfonts}%
\usepackage{amsthm}%
\usepackage{mathrsfs}%
\usepackage[title]{appendix}%
\usepackage{xcolor}%
\usepackage{textcomp}%
\usepackage{booktabs}%
\usepackage{algorithm}%
\usepackage{algorithmicx}%
\usepackage{algpseudocode}%
\usepackage{listings}%
\usepackage{bm}
\def\pmbm#1{\mbox{\boldmath $#1$}}
\usepackage{mathtools,slashed}

\newcommand\nohatslashed[1]{{{\slashed{#1}}}}
\def\mydmUN{\nohatslashed{\cal R}}

\def\opf{\Omega}

\def\ie{{\it i.e.\/}}
\def\eg{{\it e.g.\/}}
\def\viz{{\it viz.\/}}

\def\myhat{\widehat}

  \newcommand{\myargs}[1]{(\myhat{Q}_{#1} ;t_{#1})}

  \def\args{(\myhat{Q} ;t )} 
\def\myGamma{S}
\def\mySmallGamma{g}
  \def\myCtrlalpha{\Gamma}

\def\Sdim{{\cal D}}
\def\myheader#1{\noindent\textbf{#1} --}

\begin{document}

\preprint{}

\title{Proximity-measurement induced random localization in quantum fluids}


\author{Pushkar~Mohile}
\author{Paul~M.~Goldbart}
\affiliation{Department of Physics and Astronomy, Stony Brook University}

\date{30 July 2025}

\begin{abstract}
Proximity measurements probe whether pairs of particles are close to one another. We consider the impact of post-selected random proximity measurements on a quantum fluid of many distinguishable particles. We show that such measurements induce random spatial localization of a fraction of the particles, and yet preserve homogeneity macroscopically. Eventually, all particles localize, with a distribution of localization lengths that saturates at a scale controlled by the typical measurement rate. 
The steady-state distribution of these lengths is governed by a familiar scaling form. 
\end{abstract}

\maketitle

\myheader{Introduction}
The work of Li, Chen and Fisher~\cite{ref:LCP-Zeno-A,ref:LCP-Zeno-B}, and much subsequent work, has explored exciting avenues in the field of measurement-induced organization in quantum systems. 
Primarily, the focus has been on detecting and characterizing entanglement transitions in a variety of settings; see, \eg, Refs.~\cite{ref:ARCMP-Fisher,ref:Skinner-MIPT}. 
Here, in contrast, 
we address systems of particles and focus on position measurements that disrupt the prevailing tendency towards spatial uniformity. 
We find that, over time, these measurements yield increasingly pervasive localization, despite not explicitly breaking translational symmetry. 

The system we consider is a quantum fluid consisting of many {\it distinguishable\/}, mutually repelling, particles~\footnote{The standard quantum fluids consist of \emph{indistinguishable} particles. The proximity measurement considered here requires the identification, and therefore the distinguishability, of pairs of particles.}. 
We use the term {\it fluid\/} to indicate states (whether mixed or pure) in which there is space- and time-translation and space-rotation 
symmetry and all fluctuation correlations are short-ranged in space and time. 
The measurements we consider are projective and address the question: is some pair of particles closer to one another than some specified distance? 
Such measurements, which we term \textit{proximity measurements}, reveal nothing about the absolute positions of the particles, instead providing information only about their relative positions. 
While they do not explicitly break spatial translational symmetry, we shall see that they do induce spontaneous translational symmetry breaking, albeit of an unusual nature. 
What makes the measurements random are the sequences of (i)~times at which they are made and (ii)~pairs of particles about which  inquiries are made. 

We focus on post-selected outcome records in which every measurement has as its outcome: Yes, the particles in question were found to be near one another. 
The present work is a step towards the analysis of less stringent -- and therefore more readily realizable -- measurement protocols. We anticipate that such protocols would yield results that are not simple extensions of those reported here.

%
Post-selected random proximity measurements, which force particle proximity, compete with interparticle repulsions, which promote spatial uniformity. 
This competition is resolved through the emergence of an unusual collective state, in which, at any instant, some or all of the particles mutually localize one another. Each particle has its own characteristic localization length, ultimately set by the interplay between these two central tendencies.  

What is the essential physical character of this state? 
Even though simultaneous observations of particle positions would -- as for a fluid -- indicate no long-ranged spatial organization, the state of the system does in fact differ {\it qualitatively\/} from that of a fluid. 
Any two runs having the same initial conditions and random measurement sequence would yield particle positions 
that are correlated singularly strongly, 
up to an overall common translation and rotation of the system. 
The imperfectness of the correlation arises from the intrinsic fluctuations ({\it i.e.\/}, the imperfectness of the localization). 
Some aspects of the strategy for capturing random localization explored here have their origins in the theory of random solidification; see, \eg, Ref.~\cite{CGZaip-1996}.
The setting of ultra-cold atoms may provide a venue for realizing the ideas explored here. 

\myheader{Bare system}
We take the system to consist of a large number $N$ of structureless, spinless, nonrelativistic particles, each of mass $m$, interacting {\it via\/} a short-ranged, repulsive, translationally invariant two-body 
potential $U$ that promotes spatial homogeneity. 
The system is contained in a large, $\Sdim$-dimensional, hypercube of volume $V$, on which periodic boundary conditions are imposed. 
Its action 
${\cal S}_{0}$ governs quantum propagation, {\it via\/} the path-integral representation of the transition matrix elements between the (unsymmetrized, normalized) initial and final simultaneous position eigenstates: 
$({\bf x}_{1}^{\rm F},\ldots,{\bf x}_{N}^{\rm F}\vert
\,{\cal U}_{0}^{T}
\vert{\bf x}_{1}^{\rm I},\ldots,{\bf x}_{N}^{\rm I})=
\int[d{\bf r}(\cdot)]\,
\exp(i\,{\cal S}_{0}/\hbar)$. 
The path integral is taken over all system histories $\{{\bf r}_{j}(\cdot)\}_{j=1}^{N}$
connecting the initial ({\it i.e.\/}, $t=0$) and final ({\it i.e.\/}, $t=T$) configurations, 
and ${\cal U}^{T}_{0}$ is the corresponding unitary time-evolution operator. 
At $t=0$ we specify the initial state {\it via\/} a density matrix ${\cal R}(0)$. 
We assume that, absent measurements, the density matrix ${\cal R}(t)$ evolves unitarily~\footnote{Note that, here, $T$ indicates a time, not a transpose.}
and is stationary:  
${\cal R}(t)=
{\cal U}_{0}^{t\phantom\dagger}\!
{\cal R}(0)\,
{\cal U}_{0}^{t\dagger}={\cal R}(0)$. 

\smallskip\myheader{Measurements, averages, and observables}
We envision the system as being subjected to $M$ measurements, one at each of the random sequence of times 
$0<t_{1}<\cdots<t_{M}<T$. 
The $m^{\rm th}$ measurement, which occurs at time $t_{m}$, concerns a randomly chosen pair of particles 
$\{j_{m},l_{m}\}$,  
and asks whether or not their instantaneous separation 
$\vert{{\bf r}_{j_{m}}-{\bf r}_{l_{m}}}\vert$ 
is smaller than some specified length ${\cal B}$. 
To formulate the measurements, we introduce a hyperspherical indicator (or box) function $B({\bf r}-{\bf r}^{\prime})$, which takes the values 1 if the particle positions ${\bf r}$ and ${\bf r}^{\prime}$ are closer than ${\cal B}$ and 0 otherwise. 
Then, measurement $m$ is implemented by the projection operator~\footnote{Due to post-selection, the complementary operator ${\cal I}-{\cal P}_{m}$ does not feature.}
\begin{equation}
    {\cal P}_{m}\!\equiv\!\!
    \int\!\! d^{\Sdim}r_{1}\cdots d^{\Sdim}r_{N}\,
    \!\vert{\bf r}_{1},\ldots,{\bf r}_{N})
    B({\bf r}_{j_{m}}\!\!-{\bf r}_{l_{m}})
    ({\bf r}_{1},\ldots,{\bf r}_{N}\vert. 
    \nonumber
\end{equation}
\begin{widetext}
With measurements, ${\cal R}(T)$ follows {\it via\/} the unitary evolution of ${\cal R}(0)$, blended with measurement projections and normalization restoration: 
${\cal R}(T)=\mydmUN(T)\,/\,{\rm Tr}\,\mydmUN(T)$, 
where $\textrm{Tr}$ denotes the trace~\footnote{Taking the trace, ${\rm Tr}\cdots$, amounts to computing $\int d^{\Sdim}r_{1}\cdots\int d^{\Sdim}r_{N}\,({\bf r}_{1},\ldots,{\bf r}_{N}\vert\cdots\vert{\bf r}_{1},\ldots,{\bf r}_{N})$.} and 

\begin{equation}
\mydmUN(T)\equiv
\overbrace{
{\cal U}_{t_{M}}^{T\phantom\dagger}
{\cal P}_{M}\,
{\cal U}_{t_{M-1}}^{t_{M}\phantom\dagger}\!
\cdots
{\cal P}_{2}\,
{\cal U}_{t_{1}}^{t_{2}\phantom\dagger}\!\!
{\cal P}_{1}\,
{\cal U}_{0}^{t_{1}\phantom\dagger}\!\!
}^{\text{propagation forward in time}}
{\cal R}(0)\,
\overbrace{
{\cal U}_{0}^{t_{1}\dagger}\,
{\cal P}_{1}\,
{\cal U}_{t_{1}}^{t_{2}\dagger}\,
{\cal P}_{2}
\cdots
{\cal U}_{t_{M-1}}^{t_{M}\dagger}
{\cal P}_{M}\,
{\cal U}_{t_{M}}^{T\dagger}
}^{\text{propagation backward in time}}\,.
\label{eq:string}
\end{equation}
Then, for a given sequence of measurements, 
expectation values of observables ${\cal O}$ at time $T$ are given by 
  $\langle{\cal O}\rangle
={\rm Tr}\big(\mydmUN(T)\,{\cal O}\big)/\,{\rm Tr}\,\mydmUN(T)$.
The random variables that define a given measurement sequence 
({\it viz\/}.,~$\{t_{m};j_{m},l_{m}\}_{m=1}^{M}$)
are called {\it quenched\/} data. 
The particle coordinates, which evolve subject to a given measurement sequence, 
are called {\it annealed\/} variables. 
\end{widetext}

To manage the quenched disorder, we average suitable quantities over it using the replica technique and the Born rule. 
The Born rule ascribes to each measurement sequence a weight given by the quantum-mechanical probability ${\rm Tr}\big(\mydmUN(T)\big)$, which ensures, \textit{inter alia}, that the system evolution is compatible with the sequence of measurements.
This weight is augmented by a factor $\mu^{M}$ and an overall normalization factor.
Then $\mu$ provides extrinsic control over the average rate (per particle) at which measurements are made.
The physical significance of this elegant strategy, introduced by Li~\textit{et al.}~\cite{ref:LCP-Zeno-A}, is that it ensures that pairs recently measured to be near one another are more likely to be measured again~\footnote{The Born rule echos the strategy introduced by Deam and Edwards \cite{ref:Deam-1975} to model the weight ascribed to a given instance of crosslinking in vulcanized rubber.}.
We focus on detecting randomly localized states {\it via\/} the Edwards-Anderson type of order parameter~\cite{ref:GG-1987}: 
\begin{equation}
N^{-1}
\sum\nolimits_{k=1}^{N}
\prod\nolimits_{a=1}^{A}\big\langle\exp({-i{\bf p}^{a}\cdot{\bf r}_{k}(T)})\big\rangle  
\quad\,\,(A\ge 2),  
\label{eq:OPdef}
\end{equation}
which is normalized to unity when all of its wave-vector arguments $\{{\bf p}^{a}\}_{a=1}^{A}$ are zero. 
Through its dependence on these arguments, this order parameter characterizes particle organization in terms of (i)~the fraction of particles that are localized about randomly located mean positions at time $T$, and (ii)~the distribution of RMS position fluctuations exhibited by the localized particles. 
We average the order parameter~\eqref{eq:OPdef} over the quenched data with a weight given by the Born rule, and denote such averages by $[\cdots]$. 
Averaging over quenched data enables us to calculate quantities that diagnose the nature of the state induced by the measurements. 
It is expected that the phenomena that we shall capture through averaging well approximate what would be observed, at least for typical instances of the quenched data.
\smallskip\myheader{Measurement-averaged effective theory}
We now compute the order parameter~(\ref{eq:OPdef}), averaged over quenched disorder. 
We eliminate the problematic normalization factor ${\rm Tr}\big(\mydmUN(T)\big)$
using the replica technique, thus arriving at the following non-random but replica-coupled effective representation: 
\begin{widetext}
\begin{equation}
\!\!\!\!
\!\!
\bigg[
\frac{1}{N}
{\sum_{k=1}^{N}\prod_{a=1}^{A}}
\Big\langle{\rm e}^{-i{\bf p}^{a}\cdot{\bf r}_{k}(T)}\Big\rangle\bigg]
=\lim_{n\to 0}
\frac{
\Big\langle\!\!\!\Big\langle
\frac{1}{N}
{\textstyle\sum\limits_{k=1}^{N}}
{\rm e}^{-i\sum\limits_{a=1}^{A}{\bf p}^{a}\cdot{\bf r}_{k}^{a}(T)}
\!\exp\frac{\mu}{2}\!\int_{0}^{T}\!\!dt\!
\sum\limits_{j=1}^{N}\sum\limits_{l=1}^{N}
\prod\limits_{\alpha=0}^{n}\!
\Big(\!
B\big({\bf r}_{j}^{\alpha}(t)-{\bf r}_{l}^{\alpha}(t)\big)
B\big({\bar{\bf r}}_{j}^{\alpha}(t)-\bar{{\bf r}}_{l}^{\alpha}(t)\big)
\!\Big)
\Big\rangle\!\!\!\Big\rangle}
{\Big\langle\!\!\!\Big\langle
\!\exp\frac{\mu}{2}\int_{0}^{T}\!\!dt\!
\sum\limits_{j=1}^{N}\sum\limits_{l=1}^{N}
\prod\limits_{\alpha=0}^{n}
\Big(\!
B\big({\bf r}_{j}^{\alpha}(t)-{\bf r}_{l}^{\alpha}(t)\big)
B\big({\bar{\bf r}}_{j}^{\alpha}(t)-\bar{{\bf r}}_{l}^{\alpha}(t)\big)
\!\Big)
\Big\rangle\!\!\!\Big\rangle}\,.
\label{eq:quotient-form}
\end{equation}
\end{widetext}
The resulting description features $1+n$ replicas of the bare fluid, coupled as a result of the measurements, with a strength given by the measurement-rate control parameter $\mu$. 
The random measurements are responsible for the (exhibited) weighting term that couples the replicas. 
The complex \lq\lq weights\rq\rq\ in the path integral resulting from the bare liquid and initial density matrix are implied by the expectation-value notation $\langle\!\!\!\langle\,\cdots\rangle\!\!\!\rangle$. 
Also implied by this notation, and evident in the replica coupling term {\it via\/} the factor 
$B\big({\bar{\bf r}}_{j}^{\alpha}-\bar{{\bf r}}_{l}^{\alpha}\big)$, is the Schwinger-Keldysh (SK) doubling of the degrees of freedom (from $\{{\bf r}_{j}^{\alpha}\}$ to the SK pairs $\{{\bf r}_{j}^{\alpha},{\bar{\bf r}}_{j}^{\alpha}\}$), which manages the time-evolution of the density matrix~\cite{ref:SK-formalism,ref:SK+BRST}. 
\begin{widetext}
We now use Eq.~(\ref{eq:quotient-form}) to determine the behavior of the order parameter within a self-consistent mean-field approximation. 
We begin by introducing the microscopic collective coordinates
\begin{equation}
\omega({\hat p};{\hat{\bar p}};t)
\equiv
\frac{1}{N}\sum_{k=1}^{N}
\exp\big({-i{\hat p}\cdot{\hat r}_{k}(t)}\big)
\exp\big({-i{\hat{\bar p}}\cdot{\hat{\bar r}}_{k}(t)}\big),
\nonumber
\end{equation}
where hats denote replicated vectors 
[{\it e.g.\/}, $\hat p\equiv ({\bf p}^{0},{\bf p}^{1},\ldots,{\bf p}^{n})$]. 
We note that, in contrast with Eq.~(\ref{eq:quotient-form}), we now allow 
all times $t$ (in the range $0$ to $T$) and 
all coordinate-replicas (including the $0^{\rm th}$) 
as well as their SK \lq\lq doubles\rlap.\rq\rq\thinspace\ 
Next, we approximate the measurement-induced interaction exhibited in Eq.~(\ref{eq:quotient-form}) {\it via\/} the Bethe-Peierls strategy~\cite{ref:Bethe-Peierls}. 
This amounts to introducing the expectation value of $\omega$, denoted $\Omega$, 
and neglecting terms quadratic in the fluctuations $\omega-\Omega$
in the interaction term. 
Within this approximation, we arrive at the self-consistency condition for the expectation value of $\omega\,$: 
\begin{equation}
\!\Omega(\hat{p}_{0};\hat{\bar p}_{0};t_{0})
\!=\!
\lim_{n\to 0}\!
\frac{\Big\langle\!\!\!\Big\langle
\omega(\hat{p}_{0};\hat{\bar p}_{0};t_{0})
\exp\frac{N^{2}\mu}{V^{2+2n}}\!\int_{0}^{t_{0}}\!dt
\sum\limits_{\hat{p},\hat{\bar p}\,\,}\!
\prod\limits_{\alpha=0}^{n}\!\!\big(B_{\textbf{p}^{\alpha}}B_{\bar{\textbf{p}}^{\alpha}}\big)
{\rm Re}\left(\Omega(\hat{p};\hat{\bar p};t)^{\ast}
              \omega(\hat{p};\hat{\bar p};t)\right)\!\!\Big\rangle\!\!\!\Big\rangle}
{\Big\langle\!\!\!\Big\langle
\phantom{\omega(\hat{p}_{0};\hat{\bar p}_{0};t_{0})}
\exp\frac{N^{2}\mu}{V^{2+2n}}\!\int_{0}^{t_{0}}\!dt
\sum\limits_{\hat{p},\hat{\bar p}\,\,}\!
\prod\limits_{\alpha=0}^{n}\!\!\big(B_{\textbf{p}^{\alpha}}B_{\bar{\textbf{p}}^{\alpha}}\big)
{\rm Re}\left(\Omega(\hat{p};\hat{\bar p};t)^{\ast}
              \omega(\hat{p};\hat{\bar p};t)\right)\!\!\Big\rangle\!\!\!\Big\rangle}\,,
\label{eq:basic-SCE}
\end{equation}
where $B_{\bf p}=\int d^{\Sdim}r\,B({\bf r})\,\exp(-i{\bf p}\cdot{\bf r})$ 
is the the Fourier transform of the box function. 
In setting the problem up, we concerned ourselves with observations made at a final time $T$. However, in adopting the dynamical mean-field scheme, it is necessary to consider the self-consistent field $\Omega$ at all times $t_{0}$, which causality renders independent of the dynamics occurring after $t_{0}\,$.
\end{widetext}


\smallskip\myheader{Emergence of random particle localization}
Next, we identify the symmetry-breaking pattern and introduce an ansatz for the ordered state. 
The theory embodied in Eq.~(\ref{eq:quotient-form}) is symmetric under \textit{replica-dependent} translations of the SK pairs: 
for every replica, 
$({\bf r}_{j}^{\alpha},\bar{{\bf r}}_{j}^{\alpha})\to
({\bf r}_{j}^{\alpha}+{\pmbm\rho}^{\alpha},\bar{{\bf r}}_{j}^{\alpha}+{\pmbm\rho}^{\alpha})$ 
for any translation ${\pmbm\rho}^{\alpha}$.
From causality (\ie, that the measurements influence the future but not the past), it follows that the exact expectation value of $\omega$ obtained using 
Eq.~(\ref{eq:quotient-form}) depends on 
$\{{\bf p}^{\alpha},{\bar{\bf p}}^{\alpha}\}_{\alpha=0}^{n}$ 
through the center-of-mass (CM)
combinations 
${\bf Q}^{\alpha}\equiv{\bf p}^{\alpha}+{\bar{\bf p}}^{\alpha}$ 
but not the relative 
combinations 
${\bf q}^{\alpha}\equiv\big({\bf p}^{\alpha}-{\bar{\bf p}}^{\alpha}\big)/2$. 
As a result, absent spontaneous symmetry-breaking,  
we obtain the self-consistent solution 
$\Omega(\hat{p};\hat{\bar p};t)=
L(\myhat{Q})\equiv
\delta_{{\myhat{Q}},{\myhat{0}}}\equiv 
\prod\nolimits_{\alpha=0}^{n}\delta_{\textbf{Q}^{\alpha},\textbf{0}}\,$,  
which we refer to as the liquid state. 
In this state, all particles remain delocalized at all times. 

Next, we look for a state in which some fraction of the particles are localized, with mean positions that are distributed randomly and uniformly in space. 
%
These correspond to states of the effective theory in which, for every localized particle, its replicas are bound to a common center. 
The centers of each of these localized particles are distributed uniformly and independently in space. 
The symmetry of $1+n$ independent translations of the replicas is spontaneously broken down to the symmetry of common translations of the replicas. 
This is reminiscent of amorphous solidification as a consequences of crosslinking in polymer systems; for a review see Ref.~\cite{CGZaip-1996}. 
In states with such localization, the order parameter 
continuously bifurcates from the liquid state to take the form: 
\begin{subequations}
\begin{equation}
\opf\args = 
\big(1-\mySmallGamma(t)\big)\,L\args+
\mySmallGamma(t)\,\myGamma\args. 
\end{equation}
Here, $\mySmallGamma(t)$ counts the fraction of particles localized at time $t$. 
$\myGamma\args$ characterizes the randomly localized fraction; 
%
we make the following ansatz for it:   
\begin{equation}
\myGamma\args = 
\delta_{\widetilde{\bf Q},{\bf 0}} 
\int_{0}^{\infty} d\xi^2\, 
p(\xi^2;t)\,\textrm{e}^{-\myhat{Q}^{2}\xi^{2}/2},
\end{equation}%
\label{eq:Ansatz}%
\end{subequations}%
where
$\widetilde{\textbf{Q}}\equiv\sum_{\alpha=0}^{n}\textbf{Q}^{\alpha}$ 
and 
$\myhat{Q}^{2}\equiv\sum_{\alpha = 0}^{n}{|\textbf{Q}^\alpha}|^2$, 
and $p(\xi^2;t)$ has the physical meaning of the probability distribution of squared localization lengthscales at time $t$. Henceforth, we omit the limits on $\xi^{2}$ integrals.
We now show that the ordered-state ansatz~\eqref{eq:Ansatz} for $\Omega$ solves the self-consistency condition~\eqref{eq:basic-SCE}. 
To do this, we expand the RHS of~\eqref{eq:basic-SCE} perturbatively in departures from the liquid state, 
thus arriving at the following form for the self-consistency condition: 
\def\myCorrel{{\cal C}}
\begin{widetext}
\begin{eqnarray}
&&
\mySmallGamma(t_0)\Big(\myGamma\myargs{0}-L\myargs{0}\Big)
= 
\sum_{m=1}^\infty 
\frac{\left(\mu N^2\right)^m}{m!} 
\int_{0}^{t_0}\!dt_1\,\mySmallGamma(t_1)
\dots 
\int_{0}^{t_0}\!dt_m\,\mySmallGamma(t_m)
\,\frac{1}{V^{1+n}}\sum\nolimits_{\myhat{Q}_1}\cdots 
\,\frac{1}{V^{1+n}}\sum\nolimits_{\myhat{Q}_m} 
\label{eq:Ansatz-solves}
\\&&
\times
\myCorrel(\myhat{Q}_{0},\ldots,\myhat{Q}_{m};t_{0},\ldots,t_{m})
\Big(\prod\nolimits_{\alpha=0}^{n}B_{{\bf Q}_{1}^{\alpha}}\Big)
\Big(\myGamma\myargs{1}-L\myargs{1}\Big)\cdots 
\Big(\prod\nolimits_{\alpha=0}^{n}B_{{\bf Q}_{m}^{\alpha}}\Big)
\Big(\myGamma\myargs{m}-L\myargs{m}\Big). 
\nonumber
\end{eqnarray}
\end{widetext}
In Eq.~\eqref{eq:Ansatz-solves}, all relevant details concerning the system at the microscopic level are embodied in the time-ordered connected correlation functions 
$\myCorrel(\myhat{Q}_{0},\ldots,\myhat{Q}_{m};t_{0},\ldots,t_{m}),$
given by 
$\langle\!\langle
\omega(\hat{p}_{0};\hat{\bar p}_{0};t_{0})\cdots
\omega(\hat{p}_{m};\hat{\bar p}_{m};t_{m})
\rangle\!\rangle_{\textrm{con}}$.
That $\myCorrel$ does indeed depend only on the CM momenta arises from our neglect of contributions to the bare-liquid path integral~\footnote{The correlation functions in Eq.~\eqref{eq:Ansatz-solves} occur with an additional weight factor $\exp\frac{N^{2}\mu}{V^{2+2n}}\int_{0}^{t_{0}}dt \sum\nolimits_{\hat{p},\hat{\bar p}}\!\prod\nolimits_{\alpha=0}^{n}(B_{\mathbf{p}^{\alpha}}B_{\bar{\mathbf{p}}^{\alpha}})\times\mathrm{Re}\bigl(L(\hat{p}+\hat{\bar p};t)^{\ast}\omega(\hat{p};\hat{\bar p};t)\bigr)$ and corresponding normalization. An additional consequence of the neglect of phase corrections is that this weight has no impact on the correlation functions, even at nonzero measurement rate.} 
associated with relative SK paths~\footnote{This neglect also leads to a simplification of the effects of the measurement box. As it ensures that $\mathbf{r}(t)=\bar{\mathbf{r}}(t)$, and as the box function obeys $B(\mathbf{r})^{2}=B(\mathbf{r})$, it ensures that $B(\mathbf{r})B(\bar{\mathbf{r}})=B(\mathbf{r})$.}.
\def\myDiCo{D}
In the spirit of self-consistent field-theory, we approximate $\myCorrel$ by the diffusive auto-correlation part~\footnote{We note that $\mathcal{C}$ vanishes when any of its replicated momentum arguments $\hat{Q}_{r}$ vanishes.}.: 
\begin{equation}
N^{-m}\,
\delta_{\!\sum\limits_{r=0}^{m}\myhat{Q}_{r},\myhat{0}}\,
\exp\!\Big(\!{
\myDiCo
\!{\textstyle
\sum\limits_{0\le r<\bar{r}\le m}}\!\!
{\myhat{Q}}_{r}\cdot{\myhat{Q}}_{\bar{r}}
\,|t_{r}-t_{\bar{r}}|}\Big). 
\label{eq:propagators}
 \end{equation}
The scaling $N^{-m}$ arises from the neglect of correlations between distinct particles. 
The wave-vector-conserving $\delta$ factor originates in the invariance of the bare liquid under independent translations of the replicas. 

Next, we focus on the key lengthscales: 
the bare-liquid correlation length $l_\textrm{min}$, 
box diameter ${\cal B}$, 
typical localization length $\xi_{\textrm{typ}}\,$, 
and system size $V^{1/\Sdim}\,$. 
We assume that in the regime of interest they obey
$l_\textrm{min}\ll {\cal B}\ll\xi\ll L\,$. 
In this regime (because ${\cal B}\ll\xi_{\textrm{typ}}$) the measurement-box form factor varies only weakly on the localization scale $1/\xi_{\textrm{typ}}\,$, 
so we may replace $B_{{\bf Q}^\alpha}$ by $B_\textbf{0}$, \textit{i.e.\/}, 
the hypervolume $v$ of the measurement box. 
As a consequence, we obtain a single effective measurement-rate parameter 
$\myCtrlalpha\equiv\mu N\left(v/V\right)^{1+n}
\xrightarrow{n\to 0}\mu\,(N/V)v$, 
in which $\mu$ is moderated by the number of particles per measurement volume. 

Returning to the task of showing that the ansatz~\eqref{eq:Ansatz} solves the self-consistency condition~\eqref{eq:Ansatz-solves}, we focus on the wave-vector summations, taking care to treat the singular behavior at $\myhat{Q}=\myhat{0}$ before evaluating the summations as integrals and then taking the replica limit. 
This treatment has the effect of renormalizing the rate parameter such that 
$\myCtrlalpha^{m}\to\myCtrlalpha^{m}\exp\big(-\myCtrlalpha\int_{0}^{t_{0}}dt'\,\mySmallGamma(t')\big)$
at order $m$ in Eq.~\eqref{eq:Ansatz-solves}. 
Then, upon canceling the factor $\delta_{{{\widetilde{\bf Q}}}_{0},{\bf 0}}\,$, we obtain a self-consistent equation for the as-yet undetermined content of the order-parameter ansatz, \textit{viz.}, 
$\mySmallGamma(t)$ and 
the Laplace transform of $p(\xi^2;t)$ with respect to $\xi^2$, \viz, $({\cal L}{p})(K^2;t)$. 
Note that the equation is expressed in terms of the replica-free variables $t_{0}$ and $K^{2}$ [$\,\equiv\vert{{\widehat{Q}}}_{0}\vert^{2}\,$]. 
Finally, inverting the Laplace transform, we obtain
\begin{widetext}
\begin{subequations}
\begin{equation}
g(\tau)\,p(u;\tau) = {\rm e}^{-\int_{0}^\tau d\tau_1 g(\tau_1)}
    \Big\{
    \sum_{m=1}^\infty
    \int_0^\tau d\tau_1\int_0^{\tau_1}d\tau_2\int_0^{\tau_{m-1}}d\tau_m\,
    g(\tau_1) \cdots g(\tau_m)\,p_m(u;\tau,\tau_1,\dots,\tau_m)
    \Big\}, 
\label{eq:SCEfinal}
\end{equation}
where we have moved to dimensionless variables, 
one for the time $\tau\equiv\myCtrlalpha t$ and 
one for the squared length 
$u=\xi^2\,\myCtrlalpha/2\myDiCo$,
and, with a slight abuse of notation, we have retained the symbol $p$ despite having rescaled its arguments. 
The intermediate distributions $p_{m}$ in Eq.~\eqref{eq:SCEfinal} are given as follows:  
the lowest-order one is  
$p_{1}(u;\tau,\tau_{1})=\int du_{1}\,p(u_{1};\tau_{1})\,
\delta\big(u-u_{1}-(\tau-\tau_{1})\big)$; 
and the higher-order ones are recursively given by 
\begin{equation}
p_{m}(u;\tau,\tau_{1},\dots,\tau_m)
=\int du_{1}\,p(u_{1};\tau_{1})
 \int du_{2}\,p_{m-1}(u_2;\tau_{1},\dots,\tau_m)\,
 \delta\big(u-u_{12}-(\tau\!-\!\tau_1)\big), 
\label{eq:IntKerHigher}
\end{equation}
\label{eq:jointones}
\end{subequations}
\end{widetext}
where $u_{12}\equiv{u_{1}\,u_{2}/(u_1+u_2)}$ and 
with the time-ordering 
$\tau\ge\tau_{1}\ge\cdots\ge \tau_{m}\,$; 
see Ref.~\footnote{For any other orderings of the time arguments of $p_{m}$ in Eq.~\eqref{eq:IntKerHigher}, the variables on the right-hand side must be appropriately permuted.}.
We see that $p_{m}$ embodies the two essential processes: 
quantum spreading 
[\textit{i.e.}, $\xi^{2}\to\xi^{2}+2\myDiCo(t_{0}-t_{1})$] and 
measurement-induced contraction [\textit{i.e.}, $(\xi_{1}^{-2},\xi_{2}^{-2})\to{\xi}_{2}^{-2}+\xi_{1}^{-2}\,$], 
strikingly reminiscent of the processes that underlie the cavity approach to vulcanized matter~\cite{ref:RCMS-cavity}. 
We now investigate 
$\mySmallGamma(\tau)$ and $p(u;\tau)$. 

\begin{figure}
\includegraphics[width=0.483\textwidth]{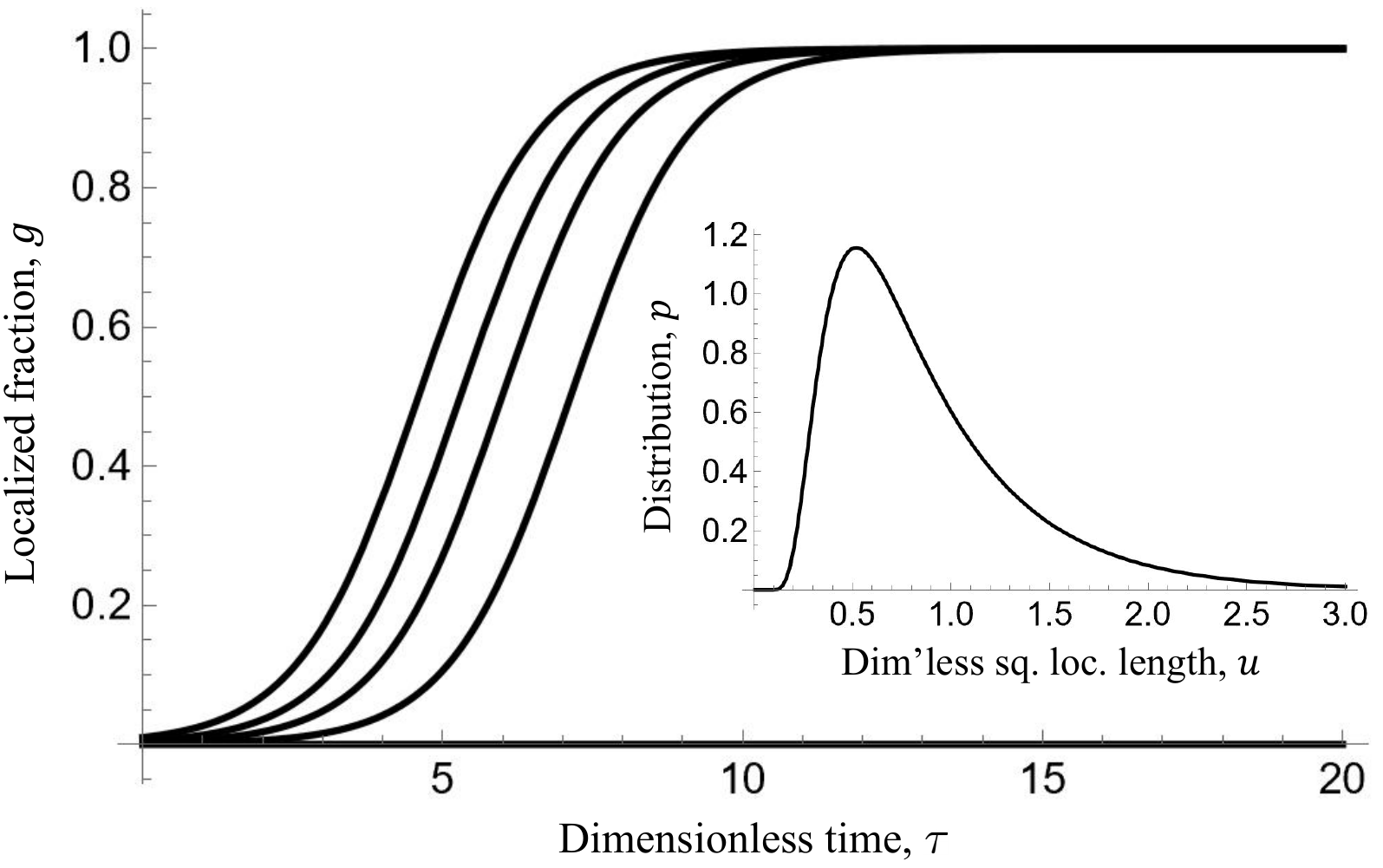}
\vskip -0.1in
\caption{\label{fig:LL-fraction}
Fraction of localized particles $g$ versus dimensionless time $\tau$. 
Left to right, $g_{0}$ is 
$0.01$, $0.005$, $0.0025$, $0.0008$, $0$. 
Inset: Scaling form of the steady-state distribution $p$ of squared localization lengths $u$. 
}
\end{figure}

The equation governing the dynamics of $\mySmallGamma$ follows from Eq.~\eqref{eq:SCEfinal} by imposing normalization on $p(u;\tau)$. It reads: 
\begin{equation}
    \mySmallGamma(\tau) = 1 - \exp\Big(-\int_{0}^{\tau}d\tau'\,\mySmallGamma(\tau')\Big).
    \label{eq:gammanocutoff}
  \end{equation}
Evidently, $\mySmallGamma(\tau)\equiv 0$ is a solution for all $\myCtrlalpha$; 
the corresponding state is the symmetry-intact, liquid state. 
It is unstable with respect to short-lived, positive-valued, symmetry-breaking perturbations. 
Such perturbations give rise to a monotonically increasing $g(\tau)$ that asymptotes to unity as $\tau\to\infty$. 
It is not surprising, given the symmetric form of the initial state and its subsequent dynamics, that a symmetry-breaking perturbation is needed to seed the growth of a state having localization. 
As a proxy for such a seed, we enforce 
$\mySmallGamma(\tau)\vert_{\tau=0}=\mySmallGamma_{0}\ne 0$; see Ref.~\footnote{In the absence of an external, symmetry-breaking perturbation, Eq.~\eqref{eq:gammanocutoff} forces $\gamma(0)=0$. Inserting a nonzero value for $\gamma(0)$ serves as a proxy for the consequences of such a perturbation.}
Then, it is straightforward to see that $\mySmallGamma(\tau)$
is the complement of the Fermi function, \ie, 
$\mySmallGamma(\tau)=
\mySmallGamma_{0}\,\exp\tau/\big((1-\mySmallGamma_{0})+\mySmallGamma_{0}\exp\tau\big),$
as shown in Fig.~\ref{fig:LL-fraction}. 
The dynamics drives the system towards $g=1$, \textit{i.e.}, ultimately all particles become localized.

We now turn to the evolution of $p(u;\tau)$. 
From Eqs.~\eqref{eq:jointones}, we find that this distribution is governed by an integral equation~\footnote{We re-sum Eq.~\eqref{eq:jointones} to obtain the all-orders dynamical equation: 
$g(\tau)\,p(u;\tau)=e^{-\int_{0}^\tau d\tau_1\,g(\tau_1)}\bigl\{\int_{0}^{\tau}d\tau_1\,g(\tau_1)\,p_{1}(u;\tau,\tau_1)
+\int_{0}^{\tau}d\tau_1\,g(\tau_1)^{2}\,e^{\int_{0}^{\tau_1}d\tau_2\,g(\tau_2)}
\int du_1\,du_2\,p(u_1;\tau_1)\,p(u_2;\tau_1)\,\delta\bigl(u-u_{12}-(\tau-\tau_1)\bigr)\bigr\}$} 
that is quadratic in $p$. 
Here, we focus on the long-time (\ie, $\tau\gg 1$) behavior, which is a steady state and thus has no $\tau$-dependence, and satisfies:
\begin{equation}
    p(u) + \frac{dp}{du} = \int du_1 \int du_2\,p(u_1)\,p(u_2)\,
    \delta\big(u-u_{12}\big). 
    \label{eq:SteadyState2}
\end{equation}
While the steady-state equation~(\ref{eq:SteadyState2}) is not solvable in terms of standard functions, its properties are well established~\cite{CGZepl-1994}: 
the distribution it gives is unimodal and peaked at $u$ of ${\cal O}(1)$, away from which it decays exponentially quickly with known asymptotics (see Fig.~\ref{fig:LL-fraction}). 
Upon restoring physical lengths, we see that the characteristic localization lengthscale in the steady state is given by 
$\myDiCo/\myCtrlalpha$, 
reflecting the interplay between the essential processes. 
This distribution is known also to govern localization lengths in vulcanized polymers~\cite{CGZepl-1994} and has arisen in other settings~\cite{ref:RBS-CBL,ref:MJS-RRN}, too.
\smallskip\myheader{Concluding remarks}
By using self-consistent replica mean-field theory, we have shown that post-selected proximity measurements progressively induce random localization of particles~\footnote{We have undertaken a similar analysis for a variant of the self-consistency condition in which the impact of a measurement is taken to be forgotten after a finite time. These conditions result in a threshold measurement rate, above which the steady state also shows random localization.}. 
In future work on this system, we anticipate investigating 
(i)~less stringent measurement protocols,
(ii)~macroscopically inhomogeneous ordering, and 
(iii)~entanglement dynamics. 
We also anticipate investigating consequences of natural analogs of proximity measurements -- particularly ones that preserve the symmetries of the underlying unitary dynamics -- in a range of quantum many-body systems, including ones composed of 
particles that are indistinguishable, particle mixtures, and spins. 

\smallskip\myheader{Acknowledgments}
We thank S.~Gopalakrishnan and A.~C.~Potter for stimulating and insightful conversations over the course of this work. 
This work was performed in part at the Aspen Center for Physics, which is supported by National Science Foundation grant PHY-2210452.


\bibliographystyle{apsrev4-2}
\bibliography{MIPL-letter-2025-07-arXiv}
\end{document}